\documentclass[aps,prb,showpacs,preprint,amssymb,amsmath,floatfix]{revtex4}
\usepackage{graphicx,bm}
\begin{document}
\bibliographystyle{prsty}
\title{Resonator design for surface electron lifetime studies using scanning tunneling spectroscopy}
\author{S. Crampin}
\affiliation{Department of Physics, University of Bath, Bath BA2 7AY,
United Kingdom}
\author{H. Jensen}
\author{J. Kr\"oger}
\author{L. Limot}
\author{R. Berndt}
\affiliation{Institut f\"ur Experimentelle und Angewandte Physik,
Christian-Albrechts-Universit\"at zu Kiel, D-24098 Kiel, Germany}
\date{\today}
\begin{abstract}
We derive expressions for the lossy boundary-scattering
contribution to the linewidth of surface electronic states
confined with atomic corrals and island resonators. Correcting
experimentally measured linewidths for these contributions along with
thermal and intrumental broadening enables
intrinsic many-body lifetimes due to electron-electron and electron-phonon
scattering to be determined. In small resonators lossy-scattering dominates
linewidths whilst different scaling of widths and separations cause levels
to merge in large resonators. Our results enable the design of
resonators suitable for lifetime studies.
\end{abstract}


\maketitle

\section{Introduction}
\label{sec:intro}

The dynamics of hot electrons and holes at surfaces have attracted
considerable attention in recent years, not least due to their
importance in photochemistry and charge-transfer processes such as
electronically-induced adsorbate reactions.
A particular focus has been the study of quasiparticle lifetimes associated
with excitations in the band of surface-localised electronic states found
at the (111) surfaces of the noble metals, which have become a testing
ground for
new theoretical, computational and experimental methods aimed
at developing a deeper fundamental understanding of quasiparticle dynamics.
\cite{ech04_}
Theoretically, much progress has been made towards a quantitative
account of inelastic electron-electron \cite{kli00a}  ($e$-$e$) and
electron-phonon \cite{eig02_} ($e$-$p$)
interactions of these states, whilst techniques such as scanning
tunneling microscopy \cite{bur99_,bra02_} (STM) and spectroscopy
\cite{jli98c} (STS), and photoelectron spectroscopy \cite{rei01_}
have been used to construct a steadily-increasing database of
experimentally-determined lifetimes.
The lifetimes of excitations with
energies spanning just a few tenths of an eV have been shown to
reflect a wealth of key surface physics; intra and interband
scattering processes; spatially-dependent and $d$-electron screening;
defect-scattering processes; electron-phonon interactions with
Rayleigh (surface) and bulk phonons; and their
temperature dependencies.

\begin{figure}[b]
\includegraphics[bbllx=79,bblly=227,bburx=533,bbury=566,
height=36mm,clip=]{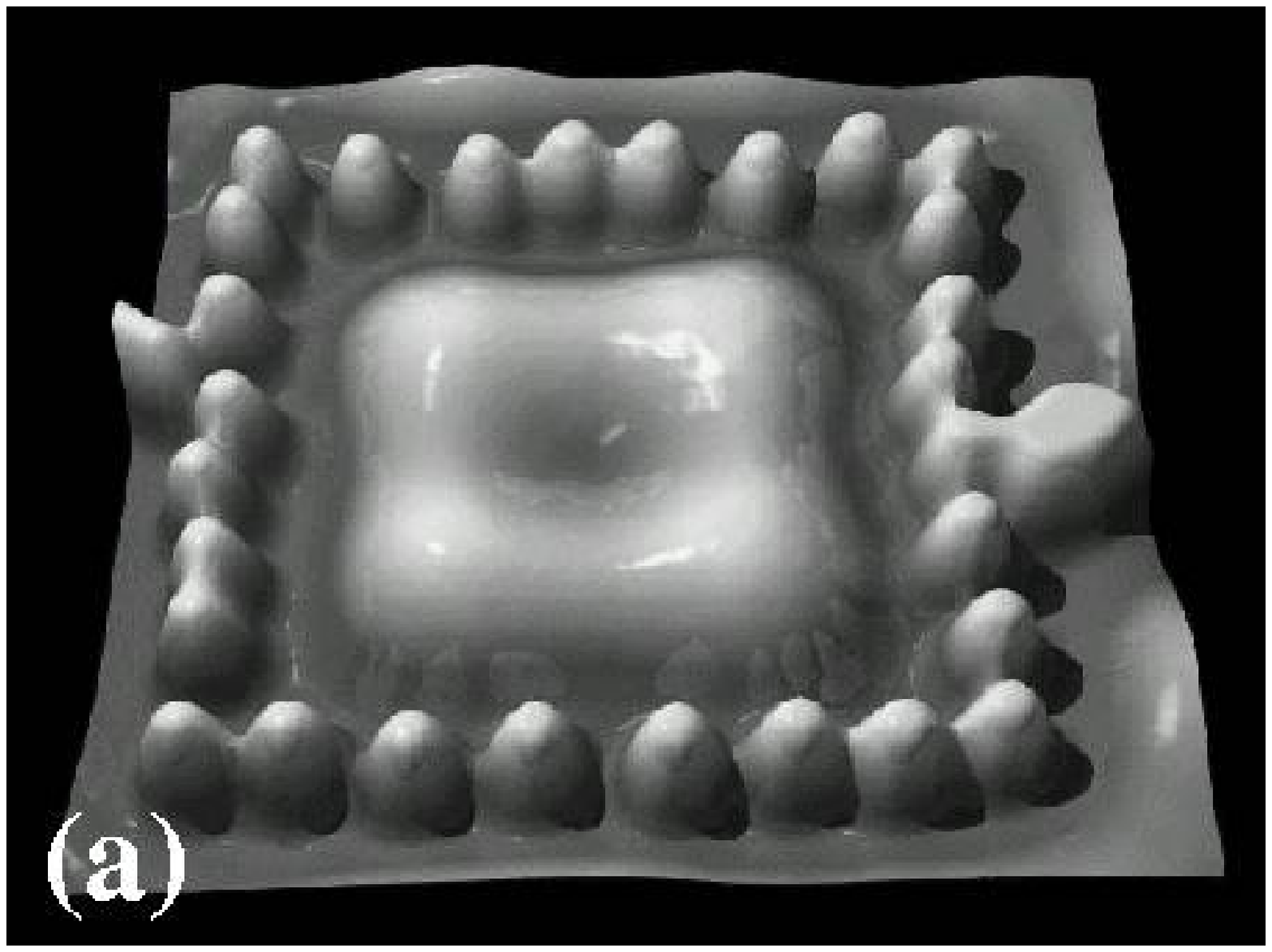}
\includegraphics[bbllx=212,bblly=304,bburx=399,bbury=488,
height=36mm,clip=]{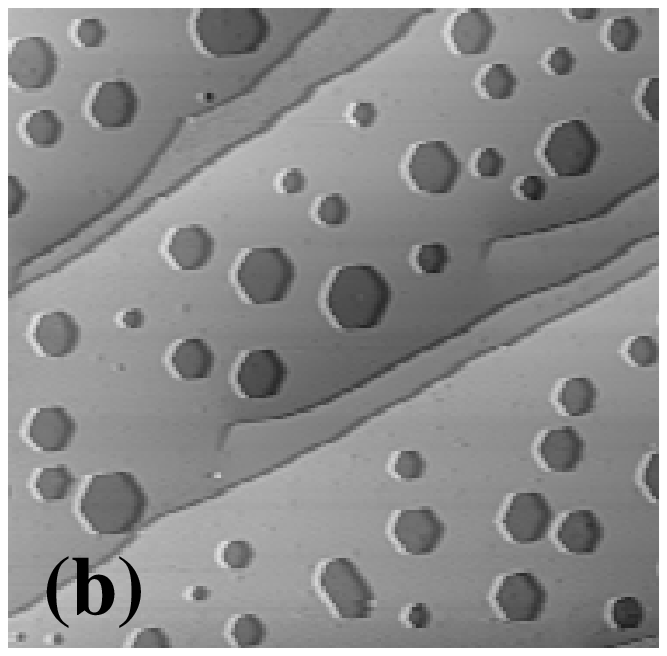}
\caption{Nanoscale electron resonators used in lifetime studies. 
(a) Pseudo three-dimensional representation of a constant current
STM image of a rectangular corral approximately 9 nm $\times$ 10 nm 
constructed from 28 Mn atoms on Ag(111). In order to enhance the atomic 
structure the intensity is plotted on a logarithmic scale.
The standing wave pattern inside is due to electron confinement \cite{kli00_}. 
(b) Greyscale representation of a constant-current STM image of a 
240 nm $\times$ 240 nm region of a Ag(111) surface showing numerous 
hexagonal monatomic vacancy islands. 
Surface state electrons are confined within these islands by strong scattering 
at the step edges of the vacancies.}
\label{fig:resonator}
\end{figure}

Lifetime studies using the STM can be divided into two approaches: those
which exploit the phase coherence of quantum interference patterns
\cite{bur99_,bra02_} and those based upon lineshape analysis \cite{jli98c}.
The lineshape method uses spectroscopic measurements of 
the differential conductivity $dI/dV$ at a fixed
position above the surface, and relies upon the
presence of spectral structure that contains a signature of the
quasiparticle lifetime. On pristine surfaces the only such feature is the
band edge onset, so that only the lifetime of excitations at this
energy are accessible. However, electron confinement to natural or artificial
nanoscale electron resonators, 
such as those shown in Figure \ref{fig:resonator},
results in energy quantisation,
inducing spectral structure in the form of
a series of resonant levels at energies that can be controlled through
hanges in the dimensions and geometry of the resonator.
The quasiparticle lifetime is then reflected in the level 
widths \cite{cra96_,kli01_},
but additional contributions arise due to lossy boundary scattering
that must be accounted for if lineshape analysis is to be used to determine
the intrinsic quasiparticle lifetime.
In this paper we provide analytic and numerical results for the linewidths
of electrons confined by different nanoscale resonators, and discuss their
use in lifetime studies using STS.

The starting point of our analyses is the equation satisfied by the
single particle Green function $G({\bf r},{\bf r}')$ \cite{economou}
\begin{equation}
\left(\! -\frac{\hbar^2\nabla^2}{2 m^\ast}\! +\!V\!-\!E\!+\!i
\text{Im}\,\Sigma_{\text{I}}\!\right)
G({\bf r},{\bf r}')
\!=\!-\delta({\bf r}\!-\!{\bf r}')
\label{eqn:g1}
\end{equation}
where ${\bf r}$ is a two-dimensional position vector,
$m^\ast$ the effective mass and  $E$ the electron
energy with respect to the surface state band minimum.  Confinement is due
to the potential $V$, which in general will remain unspecified. Only the
scattering properties of $V$ are relevant.  We also include an inelastic
potential or self energy $i\text{Im}\,\Sigma_{\text{I}}(E)$
to account for the effects of $e$-$e$ and $e$-$p$ scattering. This is related
to the lifetime $\tau_{\,\text{I}}$ associated with these processes by
$\tau_{\,\text{I}}=-\hbar/(2\text{Im}\,\Sigma_{\text{I}})$. \cite{ech04_}
The local density of surface states (LDOS) is obtained from the Green
function as
$\rho({\bf r};E)=-(2/\pi)\text{Im}G({\bf r},{\bf r})$ (the 2 is for
spin-degeneracy).  We associate the LDOS with $dI/dV$ in the usual
way \cite{ter85_}. In section \ref{sec:corral} we consider circular 
atomic corral resonators, and in section \ref{sec:islands}
circular adatom and vacancy islands. Non-circular resonators are
discussed in section \ref{sec:noncirc}. Contributions to
experimentally measured linewidths arising from thermal broadening and 
instrumental effects are described in section \ref{sec:experiment},
and in section \ref{sec:discuss} we illustrate how the results of our analyses
may be used to identify appropriate resonator structures for lifetime studies.

\section{Atomic corrals}
\label{sec:corral}

Atomic corrals are artificial structures in which individual adatoms are
positioned with atomic scale precision into closed geometries
\cite{cro93b,kli00b,kli01_,bra02_}. Strong scattering of the surface state
electrons by the surrounding adatoms leads to confinement.
The ``standard model'' for describing these systems is the
$s$-wave scattering model \cite{hel94_}, in which the solution to
(\ref{eqn:g1}) is given by \cite{hel94_,kli01_}
\begin{subequations}
\label{eqn:gms}
\begin{eqnarray}
\!\!\!\!\!\!\!\!\!G({\bm r},{\bm r}')&\!=\!&G_0({\bm r},{\bm r}')\!+\!\sum_{j,k}
G_0({\bm r},{\bm R}_j)T^{jk}G_0({\bm R}_k,{\bm r}')
\label{eqn:g0pg0Tg0}\\
T^{jk}&\!=\!&t \delta_{jk}+t\sum_{l\ne j}G_0({\bm R}_j,{\bm R}_l)T^{lk}.
\label{eqn:eom}
\end{eqnarray}
\label{eqn:sm}
\end{subequations}
The sums in (\ref{eqn:gms}) are over the
$N$ identical adatoms at locations ${\bf R}_j$, $j=0,1,..N-1$, which are
characterised by the scattering $t$-matrix
$t=(i/m^\ast)(\exp 2i\delta -1)$, where $\delta$ is the phaseshift.
$G_0$ denotes the
free-electron propagator \cite{economou}
\begin{equation}
G_0({\bm r},{\bm r}')=-\frac{im^\ast}{2\hbar^2}
H_0^{(1)}(\kappa|{\bm r}-{\bm r}'|)
\label{eqn:g0}
\end{equation}
where $H_0^{(1)}$ is a Hankel function of the first kind \cite{abram} and
$\kappa=\sqrt{2m^\ast (E-i\text{Im}\Sigma_{\text{I}})/\hbar^2}$.
In the case of a circular corral, and taking
${\bf R}_j=S(\cos \vartheta_j,\sin\vartheta_j)$, $\vartheta_j=2\pi j/N$,
when both ${\bf r}$ and ${\bf r}'$ are at the center of the corral
\begin{equation}
G-G_0=\left(-\frac{im^\ast}{2\hbar^2}
H_0^{(1)}(\kappa S)\right)^2 \frac{N}{t^{-1}-{\cal G}}
\end{equation}
and the LDOS at the corral center is
\begin{equation}
\rho=\rho_0+\frac{2N}{\pi}\left(\frac{m^\ast}{2\hbar^2}\right)^2
\text{Im}\frac{[H_0^{(1)}(\kappa S)]^2} {t^{-1}-{\cal G}}
\label{eqn:ldos}
\end{equation}
where $\rho_0=(m^\ast/\pi\hbar^2)(1\!-\!(2/\pi)\arg \kappa)$ is the clean
surface LDOS \cite{jli98c} and
\begin{subequations}
\begin{eqnarray}
{\cal G}&=&-\frac{im^\ast}{2\hbar^2}
\sum_{j=1}^{N-1}H_0^{(1)}(\kappa|{\bf R}_j-{\bf R}_0|)\\
&\approx&-\frac{im^\ast}{2\hbar^2}
\left[ NJ_0(\kappa S)H_0^{(1)}(\kappa S)-{\cal C} \right].
\end{eqnarray}
\end{subequations}
The last result follows using $|{\bf R}_j-{\bf R}_0|=2S\sin\vartheta_j/2$ and
replacing the sum by an integral over $\pi/N\le \vartheta \le 2\pi-\pi/N$.
$J_0$ is a Bessel function \cite{abram} and
${\cal C}=1+(2i/\pi)(\ln (\kappa /4 \mu)+\gamma-1)$, with
$\gamma=0.57721\ 56649...$ (Euler's constant). We have
introduced $\mu=N/2\pi S$, the linear density of atoms making up the corral.
Poles in the Green function which correspond to bound electron states
that have amplitude at the center of the corral
occur when $t^{-1}={\cal G}$. Using the asymptotic forms \cite{abram}
for $J_0$, $H_0^{(1)}$ gives the condition as
\begin{subequations}
\label{eqn:cond}
\begin{align}
&e^{2i\kappa S -i\pi/2}+1=\Omega(\kappa)
\label{eqn:cond1} \\
&\Omega(\kappa)=\frac{\kappa}{2\mu}
\left[ {\cal C}+\left(-\frac{im^\ast t}{2\hbar^2}\right)^{-1}\right].
\label{eqn:omega}
\end{align}
\end{subequations}
In the limit $\mu\rightarrow\infty$
the solutions to (\ref{eqn:cond}) are when $\kappa=\kappa_n=(n\pi-\pi/4)/S$,
corresponding to a series of resonances at energies
\begin{equation}
E_n=\frac{\hbar^2(n\pi-\pi/4)^2}{2m^\ast S^2}
\label{eqn:levels}
\end{equation}
and with widths (full width at half maximum)
\begin{equation}
\Gamma_{\text{I}n}= -2\text{Im}\Sigma_I(E_n) = \hbar/\tau_I(E_n).
\label{eqn:widths}
\end{equation}
This is the ``hard-wall'' limit, where the surface state wavefunction vanishes
at the radius of the corral. \cite{cro93b}
The exact energies in this limit have the same form as
(\ref{eqn:levels}) but with $(n\pi-\pi/4)$ ($=2.36, 5.50, 8.64$
for $n\!=\!1,2,3$ ) replaced by
the n'th zero \cite{abram} of the Bessel function $J_0$,
$j_{0,n}$ ($=2.40, 5.52, 8.65$ for
$n=1,2,3$).
This close agreement validates the use of asymptotic forms in
obtaining (\ref{eqn:cond}).

\begin{figure}[t]
\includegraphics[bbllx=182,bblly=302,bburx=398,bbury=466,
width=85mm,clip=]{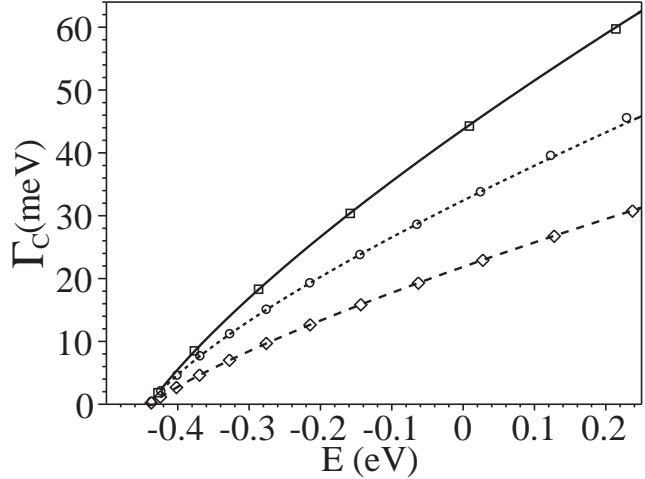}
\caption{Calculated level widths for Fe corrals on Cu(111)
using $\text{Im}\Sigma_{\text{I}}\!=\!0,
m^\ast\!=\!0.38\, m_{\text{e}}, im^\ast t\!=\!1, E_0\!=\!-0.44$\ eV.
\cite{hel94_}
Squares and lozenges denote values found solving (\ref{eqn:cond})
by Newton-Raphson iteration for $N=48, S=71.3$ \AA\
and $N=96, S=142.6$ \AA\ respectively. Circles
denote values found by non-linear least-square fitting of
a series of Lorenzians to the LDOS given
by (\ref{eqn:ldos}) for $N=48, S=142.6$ \AA.
The lines are plots of $(\hbar^2\kappa/m^\ast S){\text{Re}}
\ln [1-\Omega(\kappa)]$,
$\kappa=\sqrt{2m^\ast(E-E_0)/\hbar^2}$ for the respective corral parameters.
}
\label{fig:corral}
\end{figure}

In the hard-wall limit (\ref{eqn:widths}) shows that
the resonance widths are directly related to the
inelastic $e$-$e$ and $e$-$p$ scattering rates at the energy of the
level. However, in practice there are physical limits to how large
the linear adatom density $\mu$ can become,
with $\mu\le d_{\text{nn}}^{-1}$ where $d_{\text{nn}}$ is the
nearest-neighbour separation of surface atoms. Consequently the
confinement is always less effective than in the hard-wall limit, 
which results in increased level widths. Solving (\ref{eqn:cond}) to
first order in this  case yields level widths
\begin{equation}
\Gamma_n= \Gamma_{\text{I}n}+\Gamma_{\text{C}n}
\label{eqn:gammatot}
\end{equation}
where
\begin{equation}
\Gamma_{\text{C}n}\approx
\frac{\hbar^2 \kappa_n}{m^\ast S}\text{Re}\ln\left[1-\Omega(\kappa_n)\right].
\label{eqn:gammac}
\end{equation}
This is the contribution to the level widths due to lossy scattering
by the corral adatoms.
Figure \ref{fig:corral} shows the typical behaviour of 
$\Gamma_{\text{C}}$, showing results
for Fe corrals on Cu(111) of various sizes and linear adatom densities. 
The accuracy of the approximations made in obtaining the expression in
(\ref{eqn:gammac})
are confirmed by the close agreement to be found between the level
widths predicted by it and with those found by fitting the 
LDOS (\ref{eqn:ldos}) with a series of Lorenzians or  
from iterative solution to (\ref{eqn:cond}).
The general trend to be observed in the level widths is that they
ncrease almost linearly
with energy but are approximately halved by simultanously doubling the radius
and the number of atoms in the boundary ring.

\section{adatom and vacancy islands}
\label{sec:islands}

In a second class of resonator confinement results from
scattering at ascending or descending steps. These are adatom or vacancy
islands. \cite{avo94_,jli98b,jli99_,jen04_} In these systems
the scattering properties of the confining step are conveniently
characterised by a reflection coefficient $R$. \cite{bur98_}

Assuming circular symmetry, with $S$ now the island radius, the
Green function can be expanded as
\begin{equation}
G({\bf r},{\bf r}';E)=\frac{1}{2\pi}\sum_M
G_M(\varrho,\varrho';E)e^{iM(\varphi-\varphi')}.
\end{equation}
Substituting in to (\ref{eqn:g1}) gives for $\varrho,\varrho'<S$
\begin{equation}
\varrho^2\frac{\partial^2G_M}{\partial\varrho^2}+
\varrho\frac{\partial G_M}{\partial\varrho}+
(\kappa^2\varrho^2-M^2)G_M=
\frac{2m^\ast}{\hbar^2}\varrho\ \delta(\varrho-\varrho').
\end{equation}
Solving by the direct method gives for $\varrho<\varrho'<S$
\begin{equation}
G_M(\varrho,\varrho';E)=\frac{i\pi m^\ast}
{\hbar^2\left[A_M-1\right]} J_M(\kappa\varrho)\Psi_M(\kappa\varrho')
\label{eqn:gm}
\end{equation}
where $J_M$ is a Bessel function and
$\Psi_M(z)=H^{(1)}_M(z)+A_MH^{(2)}_M(z)$, a linear combination of
Hankel functions. The
coefficient $A_M$ is chosen to ensure that $\Psi_M$ satisfies the scattering
boundary conditions at $S$.

Identifying the energy levels from the poles in the Green function,
we see from (\ref{eqn:gm}) that these arise whenever $A_M=1$.
Only circularly symmetric states contribute to the local density of states
at the center of the island ($J_M(0)=0$ for $M\neq 0$)
so that the energy levels that can be seen by STS measurements
at the center of islands occur when $A_0=1$.
Using the asymptotic forms for the Hankel functions
\begin{eqnarray}
\Psi_0(\kappa\varrho)&\sim& \sqrt{\frac{2}{\pi \kappa\varrho}}\left[
e^{i(\kappa\varrho-\pi/4)}+A_0 e^{-i(\kappa\varrho-\pi/4)}\right]\nonumber\\
&\propto&\left[e^{i\kappa(\varrho-S)}+R e^{-i\kappa(\varrho-S)}\right],
\label{eqn:asympw}
\end{eqnarray}
that is, one can recognise within $\Psi_0$
waves incident and reflected from the confining potential, enabling the
coefficient $A_0$
to be related to the planar reflection coefficient $R$
of the step at the island edge.
Using this relationship and equating $A_0$ to 1 gives
\begin{equation}
Re^{2i\kappa S}e^{-i\pi/2}-1=0
\label{eqn:phase}
\end{equation}
as the condition for bound states visible in STS at the center of
islands. Assuming confinement sufficient to give an identifiable
series of resonant levels, and writing
$R=|R|\exp i\phi_R$, (\ref{eqn:phase}) predicts their energies as
\begin{equation}
E_n=\frac{\hbar^2(n\pi+\pi/4-\phi_R/2)^2}{2m^\ast S^2}
\label{eqn:levels2}
\end{equation}
and corresponding widths
\begin{equation}
\Gamma_n=\Gamma_{\text{I}n}+\Gamma_{\text{R}n}
\label{eqn:gamma}
\end{equation}
where
\begin{equation}
\Gamma_{\text{R}n}\approx -\frac{\hbar^2}{m^\ast}
\sqrt{\frac{2m^\ast E_n}{\hbar^2}}
\frac{\ln |R|}{S}
\label{eqn:gammar}
\end{equation}
The ``hard-wall'' limit in this case is $R\rightarrow -1$,
which ensures that the wavefunction (\ref{eqn:asympw}) vanishes at 
$S$. In this limit
(\ref{eqn:levels2}) coincides with (\ref{eqn:levels}) and
$\Gamma_{\text{R}n}=0$
so that once again the resonance widths are directly related to the
inelastic $e$-$e$ and $e$-$p$ scattering rates.
\begin{figure}[t]
\includegraphics[bbllx=177,bblly=301,bburx=398,bbury=468,
width=85mm,clip=]{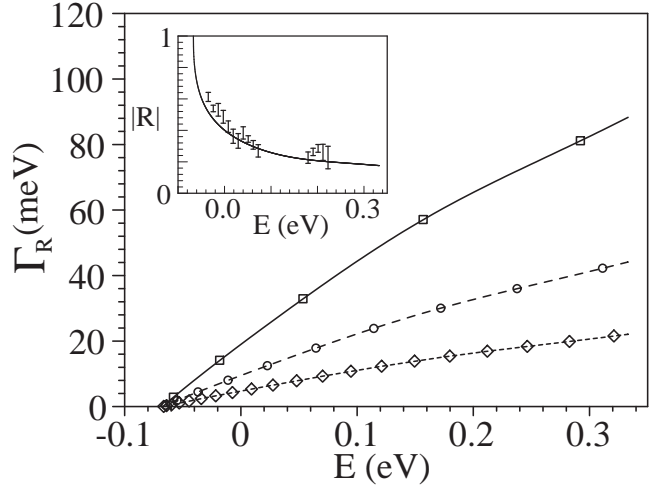}
\caption{Calculated level widths for circular Ag vacancy islands
using $\text{Im}\Sigma_I=0$, $m^\ast=0.42$, and $E_0=-67$ meV, for island radii
75 \AA (squares), 150 \AA (circles), and 300 \AA (lozenges). The lines are
plots of $-(\hbar^2/m^\ast S)\sqrt{2m^\ast (E-E_0)/\hbar^2}\ln |R|$ for the
respective island radii.
The inset shows the reflection coefficient used in the calculation (line),
along with the experimental values from Ref. [\onlinecite{bur98_}].}
\label{fig:island}
\end{figure}
In reality, scattering at real steps has been found to be lossy, 
so that $|R|<1$, and then (\ref{eqn:gammar})
enables to be estimated the resulting additional contribution to the level width
$\Gamma_{\text{R}}$. Figure \ref{fig:island} shows widths
found by solving (\ref{eqn:phase}) for variously sized Ag vacancy islands
using an energy-dependent step reflection coefficient fitted to
experimentally determined values \cite{bur98_}, along with the
widths expected using the approximate expression
(\ref{eqn:gammar}). The agreement is again very good.
The widths show similar behaviour to those in the atomic corrals,
increasing approximately linearly with energy and inversely proportional
to the island radius.

\section{non-circular resonators}
\label{sec:noncirc}
Equations (\ref{eqn:gammac}) and (\ref{eqn:gammar}) enable the 
ready determination
of the lossy scattering contribution to the spectral linewidths of electron 
states in circular resonators. In terms of lifetime studies, circular 
resonators have two distinct advantages. Firstly, the high symmetry results 
in the sparsest spectrum, which simplifies the extraction of linewidths. 
This point is discussed more fully below.  Secondly, it is not generally 
possible to obtain an analytic expression for the spectral linewidths in 
the case of non-circular resonators. Nevertheless, non-circular resonators
are important, for example the atomic structure of the (111) surfaces 
favours the natural formation of triangular and hexagonal resonators.

\begin{figure}[t]
\includegraphics[bbllx=182,bblly=303,bburx=398,bbury=466,
width=85mm,clip=]{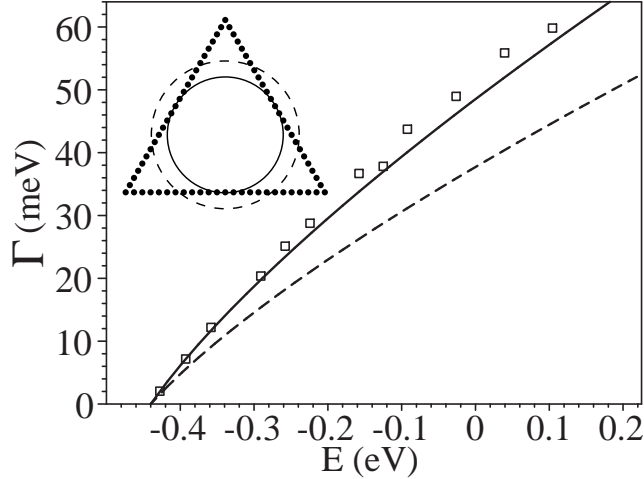}
\caption{Calculated level widths (squares) found by non-linear least-squares
fitting of a series of Lorenzians to the local density of states at the center 
of a triangular corral, side 220.5 \AA\ constructed from 72 Fe atoms 
on Cu(111) (the structure reported
in Ref. [\onlinecite{cro95_}]). Calculation parameters:  
$\text{Im}\Sigma_I=0$, $m^\ast=0.38$, $im^\ast t=1$, $E_0=-0.44$ eV. 
The lines are plots of 
the circular corral linewidth relationship, equation (\ref{eqn:gammac}),
for: dashed line, radius 81.9 \AA; and solid line, radius 63.7 \AA. 
The inset shows the atomic structure of the 
triangular corral, and the size of the two circles.}
\label{fig:triangle} 
\end{figure}

Figure \ref{fig:triangle} shows the calculated linewidths due to
lossy boundary scattering of electron
states with amplitude at the center 
of a 72 Fe atom triangular corral on Cu(111), side $d=220.5$ \AA. 
An analytic treatment for this geometry is not possible, and so
these widths have been obtained 
by fitting a series of Lorenzians to the local density of states which is 
obtained by numerically evaluating the Green function using equations 
\ref{eqn:g0pg0Tg0},\ref{eqn:eom}. 
The resonant levels occur at energies that are close to those of the hard 
wall limit,
$E_{p,q}=8\pi^2\hbar^2(p^2+q^2+pq)/(3m^\ast d^2)$ 
where $p=1,2,3,\dots$, $0\le q \le p-1$, and $p-q\ne 3\ \times$ 
integer.\cite{wli85_} 
Also shown in the figure is the linewidth relation (equation (\ref{eqn:gammac}))
for \emph{circular} corrals with radii $S=81.9$ \AA, corresponding to the
same area as the triangular corral, and $S=63.7$ \AA, corresponding to the 
largest enclosed circle. In each case in evaluating (\ref{eqn:omega})
we have used for $\mu$, the 
linear density of corral atoms, the value from the actual triangular corral,
$\mu=72/(3d)$.
The widths in the same-area circular corral underestimate the triangular
corral widths by some $\approx 25\%$, but those for the 
inscribed circular corral provide a good description.
Calculations on other
hexagonal and rectangular resonators confirm this result. The lossy
scattering contribution to the level widths at the center of 
non-circular resonators are
approximately given by the appropriate circular resonator width
relation (\ref{eqn:gammac}) or (\ref{eqn:gammar}), using for the radius that 
of the largest enclosed circle. Therefore when \emph{designing} potential
resonator structures for lifetime studies, lossy-boundary scattering effects
can be estimated using these expressions, although
more detailed calculations taking into account the actual geometry
are needed for subsequent \emph{quantitative} analysis of experimental spectra.

\section{Thermal and instrumental effects}
\label{sec:experiment}

\begin{figure}[t]
\includegraphics[bbllx=237,bblly=311,bburx=519,bbury=464,
width=85mm,clip=]{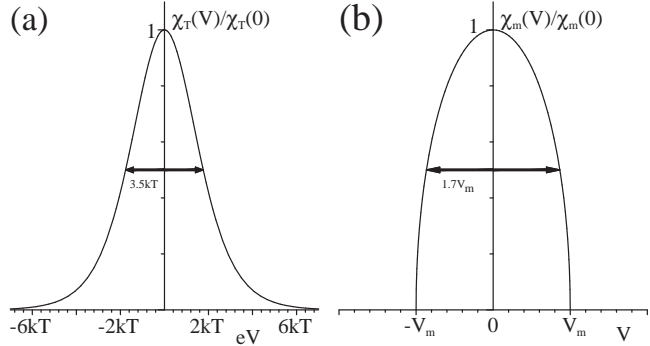}
\caption{(a) Thermal broadening function, resulting from the temperature 
dependence of the occupation of electron states. 
(b) Instrumental broadening associated with the use of a modulating voltage
and lock-in technique for recording the differential conductivity.}
\label{fig:broaden}
\end{figure}

For completeness, we note that in addition to the lossy-scattering contribution,
the experimental linewidth also includes contributions due to
instrumental and thermal effects. These must also be borne in mind when
considering the use of resonators for lifetime determinations.
The thermal broadening is induced by the temperature dependence of the
Fermi-Dirac distribution function,\cite{ter85_} which 
enters via the requirement that electrons tunnel in STS from occupied states
in the tip/sample to unoccupied states in the sample/tip.
The effect on the differential
conductivity is to convolute with the derivative of the Fermi-Dirac 
distribution, 
\begin{equation}
\chi_T(V)=\chi_T(0)\cosh^{-2}(\text{e}V/2\text{k}T).
\label{eqn:chit}
\end{equation}
The thermal broadening function shown in Figure \ref{fig:broaden}
has a FWHM of $3.5\text{k} T$. This is a relatively 
small contribution at low temperatures (=$1.5\,\text{mV}$ at $T=10\,\text{K}$).

The differential conductivity is normally measured using a lock-in technique,
which reduces phase-incoherent noise contributions. \cite{jkl_73}
A sinusoidal voltage modulation $V_{\text{m}}\cos(\omega t)$
is superimposed upon the tunneling voltage $V$,
and the signal at frequency $\omega$ recorded:
\begin{eqnarray}
I_\omega(V)&=&
\frac{2\omega}{\pi V_{\text{m}}}
\int_0^{\pi/\omega}I(V+V_{\text{m}}\cos\omega t)\cos\omega t\,dt
\nonumber \\
&=&
\int_{-\infty}^\infty\frac{dI}{dV}(V+V')\chi_{\text{m}}(V')\,dV'.
\end{eqnarray}
The measured voltage is the differential conductivity convoluted with the
instrumental function $\chi_{\text{m}}$, 
\begin{equation}
\chi_{\text{m}}(V)=\left\{
\begin{array}{ll}
2\sqrt{V_{\text{m}}^2-V^2}/\pi V_{\text{m}}^2\qquad
& \left|V\right|\le V_{\text{m}} \\
0 & \left|V\right|> V_{\text{m}}.
\end{array}
\right.
\label{eqn:chim}
\end{equation}
The FWHM of this instrumental broadening function $\chi_{\text{m}}$, 
shown in Figure \ref{fig:broaden}, is $1.7 V_{\text{m}}$. 

\section{discussion}
\label{sec:discuss}

The results of the previous sections permit
the design and use of resonator structures for the measurement of
intrinsic lifetimes.
To illustrate this, we consider circular atomic corrals constructed from
\begin{figure}[t]
\includegraphics[bbllx=178,bblly=301,bburx=405,bbury=468,
width=85mm,clip=]{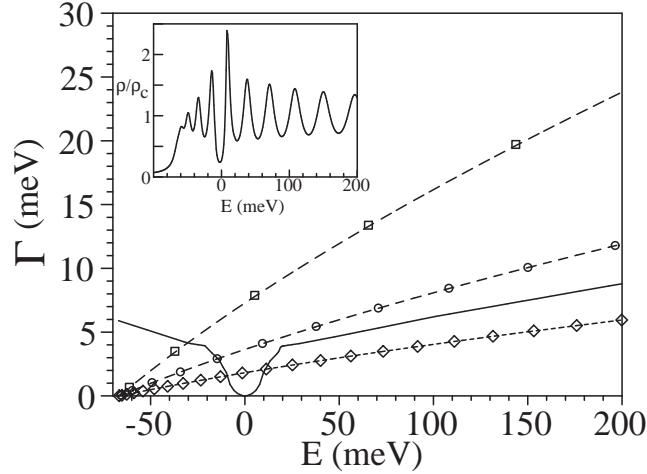}
\caption{Linewidth contributions to states at the center of
circular Ag adatom corrals on Ag(111). Intrinsic width due to $e$-$e$ and
$e$-$p$ scattering (solid line --- compiled from theoretical values reported
in Refs. [\onlinecite{kli00a,vit03_,eig02_}]). Lossy boundary scattering:
60 atoms corral, radius 100 \AA (squares); 120 atoms, radius 200 \AA (circles);
240 atoms, radius 300 \AA (lozenges).  Inset: the LDOS
normalised by $\rho_{\text{C}}=m^\ast/\pi\hbar^2$ at the center of the
120 atom corral.}
\label{fig:agcorral}
\end{figure}
Ag adatoms on Ag(111), for which a scattering phaseshift
$\delta=0.75+0.42i$ has been measured. \cite{bra02_}
The basic idea is that resonators with different dimensions result in
different series of energy levels, at energies given by (\ref{eqn:levels})
(or more accurately by numerically
solving (\ref{eqn:cond})). By choosing an
appropriate radius $S$,
levels can be positioned at specific energies $\epsilon=E_n$.
Using STS to measure the corresponding
level width $\Gamma$, equations (\ref{eqn:gammatot})
and (\ref{eqn:gammac}) plus knowledge of the thermal and instrumental
broadening effects (section \ref{sec:experiment}) may then
be used to identify $\tau_{\text{I}}(\epsilon)=\hbar/\Gamma_{\text{I}n}$.

Figure \ref{fig:agcorral} compares theoretical estimates for the 
intrinsic many-body widths of the Ag(111) Shockley surface state
with the boundary-loss contributions $\Gamma_{\text{C}n}$ that arise in
variously-sized circular corrals.
For the smallest corral of radius $S=10$ nm $\Gamma_{\text{C}}$ (squares)
is significantly greater than $\Gamma_{\text{I}}$ (solid line) for 
most energies.
In practice this will make harder
the accurate determination of $\Gamma_{\text{I}}$ and
hence lifetimes $\tau_I$ ($=\hbar/\Gamma_{\text{I}}$)
from STS-measured linewidths, which will be the sum
of $\Gamma_{\text{I}}$ and $\Gamma_{\text{C}}$, plus the effects of the
convolution
with $\chi_T$ (\ref{eqn:chit}) and $\chi_{\text{m}}$ (\ref{eqn:chim}).
In the 20 nm radius corral the
intrinsic and boundary-loss linewidth contributions are comparable for most
energies, making it a better candidate structure for intrinsic lifetime 
determinations.
One might consider going further, noting that for a fixed linear density 
of atoms making up the corral, $\mu=N/(2\pi S)$, the linewidth 
$\Gamma_{\text{C}}$ decreases linearly with $S$, encouraging 
the use of corrals of even greater radius. For almost all energies
the lossy-scattering contribution
$\Gamma_{\text{C}}$
in the 30 nm corral shown in Fig. \ref{fig:agcorral} is smaller than
the intrinsic linewidth
$\Gamma_{\text{I}}$.
However, from (\ref{eqn:levels}) we see that the level spacings vary as
$1/S^2$, decreasing with radius more rapidly than
$\Gamma_{\text{C}}$, so that in larger corrals the resonances merge,
and the levels lose their integrity. 
This is already apparent at low energies in the 20 nm corral. The LDOS for
this case is presented in the inset in Fig. \ref{fig:agcorral}, and shows
that the lowest two levels are only just separated. If typical thermal and 
instrumental broadenings are included, then the lowest level becomes 
an indiscernable shoulder.
Hence lifetimes
$\tau_{\text{I}}^{\text{Ag}}$
are best measured
in Ag adatom corrals with radii up to $\sim 20$ nm, with
smaller resonators used
for lifetimes at energies towards the bottom of the band of
surface states.

We conclude by summarising the main results of this work. 
We have derived analytic expressions for the contribution 
to the spectral linewidths of electronic states measured using
scanning tunneling spectroscopy at the 
centre of circular atomic corrals and adatom and vacancy islands
that arises due to lossy boundary-scattering.
Correcting measured linewidths for these contributions as well as
thermal and instrumental broadening effects enables
intrinsic linetimes due to electron-electron and electron-phonon scattering 
to be determined. The expressions that we have obtained 
are straightforward to evaluate, and 
also provide an estimate of lossy-confinement effects in non-circular 
resonators where the relevant radius is that of the largest enclosed 
circle.  Using our linewidth expressions it is possible to design 
resonator structures appropriate for lifetime studies.

\section*{Acknowledgments}
S. C. acknowledges the support of the British Council.
H. J., J. K., L. L. and R. B. thank the Deutsche Forschungsgemeinschaft for
financial support.

\end{document}